\documentclass[ALICE,manyauthors]{cernphprep}
\usepackage[comma,square,numbers,sort&compress]{natbib}
\usepackage{hyperref}
\usepackage{lineno}
\usepackage{xspace}
\usepackage{xcolor}
 \usepackage{upgreek}
 \usepackage{booktabs}
 \usepackage{appendix}
 
 \usepackage[T1]{fontenc}
\usepackage{orcidlink} 

\begin{document}
%

\newcommand{\pp}           {pp\xspace}
\newcommand{\ppbar}        {\mbox{$\mathrm {p\overline{p}}$}\xspace}
\newcommand{\XeXe}         {\mbox{Xe--Xe}\xspace}
\newcommand{\PbPb}         {\mbox{Pb--Pb}\xspace}
\newcommand{\pA}           {\mbox{pA}\xspace}
\newcommand{\pPb}          {\mbox{p--Pb}\xspace}
\newcommand{\AuAu}         {\mbox{Au--Au}\xspace}
\newcommand{\dAu}          {\mbox{d--Au}\xspace}

\newcommand{\s}            {\ensuremath{\sqrt{s}}\xspace}
\newcommand{\sNN}          {\ensuremath{\sqrt{s_{\mathrm{NN}}}}\xspace}
\newcommand{\pt}           {\ensuremath{p_{\rm T}}\xspace}
\newcommand{\kEt}          {\ensuremath{kE_{\text{T}}\xspace}}
\newcommand{\kEtncq}       {\ensuremath{kE_{\text{T}}/n_{\mathrm{q}}}\xspace}
\newcommand{\meanpt}       {$\langle p_{\mathrm{T}}\rangle$\xspace}
\newcommand{\ycms}         {\ensuremath{y_{\rm CMS}}\xspace}
\newcommand{\ylab}         {\ensuremath{y_{\rm lab}}\xspace}
\newcommand{\etarange}[1]  {\mbox{$\left | \eta \right |~<~#1$}}
\newcommand{\yrange}[1]    {\mbox{$\left | y \right |~<~#1$}}
\newcommand{\dndy}         {\ensuremath{\mathrm{d}N_\mathrm{ch}/\mathrm{d}y}\xspace}
\newcommand{\dndeta}       {\ensuremath{\mathrm{d}N_\mathrm{ch}/\mathrm{d}\eta}\xspace}
\newcommand{\avdndeta}     {\ensuremath{\langle\dndeta\rangle}\xspace}
\newcommand{\dNdy}         {\ensuremath{\mathrm{d}N_\mathrm{ch}/\mathrm{d}y}\xspace}
\newcommand{\Npart}        {\ensuremath{N_\mathrm{part}}\xspace}
\newcommand{\Ncoll}        {\ensuremath{N_\mathrm{coll}}\xspace}
\newcommand{\dEdx}         {\ensuremath{\textrm{d}E/\textrm{d}x}\xspace}
\newcommand{\RpPb}         {\ensuremath{R_{\rm pPb}}\xspace}
\newcommand{\RAA}          {\ensuremath{R_{\rm AA}}\xspace}
\newcommand{\QpPb}         {\ensuremath{Q_{\rm pPb}}\xspace}
\newcommand{\Qcp}          {\ensuremath{Q_{\rm CP}}\xspace}
\newcommand{\vn}           {\ensuremath{v_{\rm n}}\xspace}
\newcommand{\vtwo}         {\ensuremath{v_{\rm 2}}\xspace}
\newcommand{\vtwoncq}     {\ensuremath{v_{\rm 2}/n_{\rm q}}\xspace}

\newcommand{\Ntrkl}        {\ensuremath{N_\mathrm{trkl}}\xspace}
\newcommand{\averNtrkl}    {\ensuremath{\langle N_\mathrm{trkl}\rangle}\xspace}
\newcommand{\Nch}          {\ensuremath{N_\mathrm{ch}}\xspace}
\newcommand{\averNch}      {\ensuremath{\langle N_\mathrm{ch}\rangle}\xspace}
\newcommand{\zVtx}         {\ensuremath{z_\mathrm{vtx}}\xspace}
\newcommand*\code[1]{\texttt{#1}}
\newcommand{\fprompt}      {\ensuremath{f_\mathrm{prompt}}\xspace}
\newcommand{\fnonprompt}   {\ensuremath{f_\mathrm{non\text{-}prompt}}\xspace}
\newcommand{\AccEff}       {\ensuremath{\mbox{$\text{Acc}\times\epsilon$}}\xspace}
\newcommand{\effNP}[1]         {\ensuremath{(\mathrm{Acc}\times\epsilon)^\mathrm{non\text{-}prompt}_{#1}}\xspace}
\newcommand{\effP}[1]          {\ensuremath{(\mathrm{Acc}\times\epsilon)^\mathrm{prompt}_\mathrm{#1}}\xspace}
\newcommand{\AccEffNP}         {\ensuremath{(\mathrm{Acc}\times\epsilon)_{\mathrm{non\text{-}prompt}}}\xspace}
\newcommand{\rawY}[1]          {\ensuremath{Y_\mathrm{#1}}\xspace}
\newcommand{\Nraw}               {\ensuremath{N_\mathrm{raw}}\xspace}
\newcommand{\fP}           {\ensuremath{f_\mathrm{p}}\xspace}
\newcommand{\fNP}          {\ensuremath{f_\mathrm{np}}\xspace}
\newcommand{\Np}               {\ensuremath{N_\mathrm{prompt}}\xspace}
\newcommand{\Nnp}              {\ensuremath{N_\mathrm{non\text{-}prompt}}\xspace}
\newcommand{\HbtoLc}       {\ensuremath{\mathrm{h_b}\to\Lc+\mathrm{X}}\xspace}
\newcommand{\HbtoHc}       {\ensuremath{\mathrm{h_b}\to\mathrm{h_c}+\mathrm{X}}\xspace}
\newcommand{\fbtoLb}       {\ensuremath{f(\mathrm{b}\to\Lb)}\xspace}
\newcommand{\fbtoHb}       {\ensuremath{f(\mathrm{b}\to\mathrm{h_b})}\xspace}
\newcommand{\fbtoB}       {\ensuremath{f(\mathrm{b}\to\mathrm{B})}\xspace}
\newcommand{\fbtoBee}      {\ensuremath{f(\mathrm{b}\to\mathrm{B})_{\mathrm{e^{+}e^{-}}}}\xspace}
\newcommand{\fbtoLbLHCb}       {\ensuremath{f(\mathrm{b}\to\Lb)_{\mathrm{LHCb}}\xspace}}

\newcommand{\LT}           {L{\'e}vy-Tsallis\xspace}
\newcommand{\TeV}          {\ensuremath{\mathrm{TeV}}\xspace}
\newcommand{\GeV}          {\ensuremath{\mathrm{GeV}}\xspace}
\newcommand{\GeVc}         {\ensuremath{\mathrm{GeV}/c}\xspace}
\newcommand{\MeV}          {\ensuremath{\mathrm{MeV}}\xspace}
\newcommand{\lumi}         {\ensuremath{\mathcal{L}}\xspace}
\newcommand{\de}           {\ensuremath{\mathrm{d}}\xspace}
\newcommand{\mub}          {\ensuremath{\upmu\mathrm{b}}\xspace}

\newcommand{\ITS}          {\rm{ITS}\xspace}
\newcommand{\TOF}          {\rm{TOF}\xspace}
\newcommand{\ZDC}          {\rm{ZDC}\xspace}
\newcommand{\ZDCs}         {\rm{ZDCs}\xspace}
\newcommand{\ZNA}          {\rm{ZNA}\xspace}
\newcommand{\ZNC}          {\rm{ZNC}\xspace}
\newcommand{\SPD}          {\rm{SPD}\xspace}
\newcommand{\SDD}          {\rm{SDD}\xspace}
\newcommand{\SSD}          {\rm{SSD}\xspace}
\newcommand{\TPC}          {\rm{TPC}\xspace}
\newcommand{\TRD}          {\rm{TRD}\xspace}
\newcommand{\VZERO}        {\rm{V0}\xspace}
\newcommand{\VZEROA}       {\rm{V0A}\xspace}
\newcommand{\VZEROC}       {\rm{V0C}\xspace}
\newcommand{\Vdecay} 	   {\ensuremath{V^{0}}\xspace}

\newcommand{\ee}                {\ensuremath{\mathrm{e^{+}e^{-}}}\xspace} 
\newcommand{\pip}               {\ensuremath{\uppi^{+}}\xspace}
\newcommand{\pim}               {\ensuremath{\uppi^{-}}\xspace}
\newcommand{\pipm}               {\ensuremath{\uppi^{\pm}}\xspace}
\newcommand{\kap}               {\ensuremath{\rm{K}^{+}}\xspace}
\newcommand{\kam}               {\ensuremath{\rm{K}^{-}}\xspace}
\newcommand{\pbar}              {\ensuremath{\rm\overline{p}}\xspace}
\newcommand{\kzero}             {\ensuremath{{\rm K}^{0}_{\rm{S}}}\xspace}
\newcommand{\lmb}               {\ensuremath{\Lambda}\xspace}
\newcommand{\almb}              {\ensuremath{\overline{\Lambda}}\xspace}
\newcommand{\Om}                {\ensuremath{\Omega^-}\xspace}
\newcommand{\Mo}                {\ensuremath{\overline{\Omega}^+}\xspace}
\newcommand{\X}                 {\ensuremath{\Xi^-}\xspace}
\newcommand{\Ix}                {\ensuremath{\overline{\Xi}^+}\xspace}
\newcommand{\Xis}               {\ensuremath{\Xi^{\pm}}\xspace}
\newcommand{\Oms}               {\ensuremath{\Omega^{\pm}}\xspace}
\newcommand{\degree}            {\ensuremath{^{\rm o}}\xspace}
\newcommand{\DzerotoKpi}        {\ensuremath{{\rm D}^0 \to {\rm K}^-\uppi^+}\xspace}
\newcommand{\DplustoKpipi}      {\ensuremath{{\rm D}^+\to {\rm K}^-\uppi^+\uppi^+}\xspace}
\newcommand{\DstartoDpi}        {\ensuremath{{\rm D}^{*+} \to {\rm D}^0 \uppi^+}\xspace}
\newcommand{\DstartoDpiPrecise} {\ensuremath{{\rm D^{*+}(2010)\to D^0\uppi^+}}\xspace}
\newcommand{\Dstophipi}         {\ensuremath{{\rm D_s^{+}\to \upphi\uppi^+}}\xspace}
\newcommand{\DstophipitoKKpi}   {\ensuremath{{\rm D_s^{+}\to \upphi\uppi^+\to K^-K^+\uppi^+}}\xspace}
\newcommand{\phitoKK}           {\ensuremath{{\rm \upphi\to  K^-K^+}}\xspace}
\newcommand{\Dzero}             {\ensuremath{\mathrm{D^0}}\xspace}
\newcommand{\Dzerobar}          {\ensuremath{\mathrm{\overline{D}^0}}\xspace}
\newcommand{\Dstar}             {\ensuremath{\mathrm{D^{*+}}}\xspace}
\newcommand{\Dstarm}            {\ensuremath{\mathrm{D^{*-}}}\xspace}
\newcommand{\Dplus}             {\ensuremath{\mathrm{D^+}}\xspace}
\newcommand{\Dminus}            {\ensuremath{\mathrm{D^-}}\xspace}
\newcommand{\Ds}                {\ensuremath{\mathrm{D_s^+}}\xspace}
\newcommand{\Lc}                {\ensuremath{\Lambda_\mathrm{c}^+}\xspace}
\newcommand{\Lcpm}              {\ensuremath{\Lambda_\mathrm{c}^\pm}\xspace}
\newcommand{\Ks}                {\ensuremath{\mathrm{K_S^0}}\xspace}
\newcommand{\LctopKzeros}       {\ensuremath{\Lambda_\mathrm{c}^+\to\mathrm{p}\Ks}\xspace}
\newcommand{\Kstopipi}          {\ensuremath{\Ks\to\uppi^+\uppi^-}\xspace}
\newcommand{\LctopKpi}          {\ensuremath{\Lambda_\mathrm{c}^+\to\mathrm{pK^-\uppi^+}}\xspace}
\newcommand{\KKpi}              {\ensuremath{{\rm K^-K^+\uppi^+}}\xspace}
\newcommand{\Bplus}             {\ensuremath{\mathrm{B^+}}\xspace}
\newcommand{\Bs}                {\ensuremath{\mathrm{B_s^0}}\xspace}
\newcommand{\Bzero}             {\ensuremath{\mathrm{B^0}}\xspace}
\newcommand{\Lb}                {\ensuremath{\Lambda_\mathrm{b}^0}\xspace}
\newcommand{\Jpsi}                {\ensuremath{\mathrm{J}/\uppsi}\xspace}
\newcommand{\bbbar}             {\ensuremath{\mathrm{b\overline{b}}}\xspace}
\newcommand{\ccbar}             {\ensuremath{\mathrm{c\overline{c}}}\xspace}
\newcommand{\Hc}                {\ensuremath{\mathrm{h_c}}\xspace}
\newcommand{\Hb}                {\ensuremath{\mathrm{h_b}}\xspace}

\newcommand{\fonll}            {\textsc{fonll}\xspace}
\newcommand{\nlo}             {\textsc{nlo}\xspace}
\newcommand{\nnlo}             {\textsc{nnlo}\xspace}
\newcommand{\powheg}           {\textsc{Powheg}\xspace}
\newcommand{\gmvfns}           {\textsc{gm-vfns}\xspace}
\newcommand{\pythia}           {\textsc{Pythia~8}\xspace}
\newcommand{\pythiaprec}       {\textsc{Pythia~8.243}\xspace}
\newcommand{\geantthree}       {\textsc{Geant3}\xspace}
\newcommand{\clrblc}           {\textsc{clr-blc}\xspace}
\newcommand{\cteq}             {\textsc{cteq6.6}\xspace}
\newcommand{\ctnloten}             {\textsc{ct10nlo}\xspace}
\newcommand{\ctnlofourteen}             {\textsc{ct14nlo}\xspace}
\newcommand{\rqm}           {\textsc{rqm}\xspace}
\newcommand{\tamu}            {\textsc{tamu}\xspace}
\newcommand{\Lint}             {\ensuremath{\mathcal{L}_\mathrm{int}}\xspace}

\begin{titlepage}
\PHyear{2026}       
\PHnumber{070}      
\PHdate{11 March}  

    \title{Evidence of different \Lc-baryon and D-meson elliptic flow in Pb--Pb collisions at $\mathbf{\sqrt{\textit{s}_\mathrm{NN}}}$ = 5.36 TeV with ALICE at the LHC}
    \ShortTitle{Charm hadrons elliptic flow in \PbPb collisions}   

    \Collaboration{ALICE Collaboration\thanks{See Appendix~\ref{app:collab} for the list of collaboration members}}
    \ShortAuthor{ALICE Collaboration} 

    \begin{abstract}
The ALICE collaboration reports the azimuthal-anisotropy coefficient \vtwo of prompt \Dzero, \Dplus, \Ds mesons and the first measurement of \vtwo of prompt \Lc baryons in semicentral \PbPb collisions at a center-of-mass energy per nucleon pair of \sNN = 5.36 TeV. The D mesons and \Lc baryons are reconstructed in their hadronic decays at midrapidity ($\yrange0.8$) in the transverse-momentum interval 0.5 $<$ \pt $<$ 24 \GeVc. Similar \vtwo values are measured for \Dzero and \Dplus, while a hint of a difference ($2.6\sigma$) emerges between \Dzero and \Ds mesons in the $1 <\pt < 5~\GeVc$ interval. A larger \vtwo for \Lc baryons with respect to \Dzero mesons is observed with $3.7\sigma$ significance for $4 < \pt < 12~\GeVc$, providing evidence for the partonic origin of charm-hadron \vtwo and hadron formation via quark coalescence. This interpretation is further supported by comparisons with theoretical calculations of charm-quark transport in a hydrodynamically expanding medium.

    \end{abstract}
\end{titlepage}

\setcounter{page}{2} 

Quantum chromodynamics predicts that, under extreme conditions of high temperature and energy density, strongly-interacting matter undergoes a transition from the hadronic phase to a color-deconfined
medium, the quark--gluon plasma (QGP)~\cite{Fukushima:2010bq,HotQCD:2018pds,Borsanyi:2020fev}. Ultrarelativistic heavy-ion collisions create the conditions for such a hot, dense medium to form and probe its properties~\cite{STAR:2005gfr,PHENIX:2004vcz,ALICE:2022wpn,CMS:2024krd}.
The QGP created in this regime behaves as a nearly perfect fluid with a shear viscosity-to-entropy-density ratio, $\eta/s$, close to the minimum value predicted by theory~\cite{Kovtun:2004de}.
Heavy quarks, i.e.\ charm and beauty, because of their large masses, are predominantly produced in initial hard-scattering processes prior to QGP formation~\cite{Liu:2012ax,Andronic:2015wma}. As a result, they traverse all stages of the system evolution, interacting with the medium constituents through both elastic (collisional)~\cite{Braaten:1991we} and inelastic (gluon radiation)~\cite{Gyulassy:1990ye,Baier:1996sk} processes, being sensitive to QGP transport properties~\cite{Prino:2016cni,Dong:2019byy}. Evidence for charm-quark in-medium interactions is provided by the strong modification of the transverse-momentum (\pt) distributions of charm hadrons in nucleus--nucleus collisions relative to proton-proton collisions, as observed at both RHIC~\cite{Adare:2010de,Abelev:2006db,Adamczyk:2014uip,Adler:2005xv,Adare:2015hla} and LHC~\cite{Adam:2015sza,Abelev:2012qh,Adam:2016khe,Sirunyan:2017xss,ALICE:2021rxa, ATLAS:2020yxw} energies. These modifications arise from momentum exchanges with the medium, with energetic quarks predominantly losing energy, resulting in a suppression of the yield of high-\pt charm hadrons.

In non-central nucleus--nucleus collisions, the initial geometrical anisotropy of the overlap region is converted, through multiple interactions among the medium constituents, into an anisotropy of the final-state hadron momentum distribution. This phenomenon, known as ``anisotropic flow'', constitutes a key observable sensitive to the properties of the medium. Anisotropic flow can be quantified by the Fourier coefficients \vn, given by $\vn =  \langle\cos[n(\varphi-{\rm \Psi_{n}})]\rangle$, where $\langle \rangle$ denotes an average over all particles in a single event, $\varphi$ is the particle momentum azimuth, and $\rm \Psi_{n}$ is the azimuth of the collision symmetry plane for the $n^{\mathrm{th}}$ harmonic~\cite{Voloshin:1994mz}. The second-order Fourier coefficient $v_{2}$, referred to as ``elliptic flow'', is the dominant flow harmonic in semicentral heavy-ion collisions. The measured sizable D-meson \vtwo~\cite{STAR:2017kkh,ALICE:2018gif,ALICE:2020iug,ALICE:2021kfc,CMS:2020bnz,CMS:2021qqk} indicates that charm quarks participate in the collective flow of the medium and reflects how the azimuthal distribution of charm-hadron momenta is influenced by the charm-quark interactions with the medium. At low momenta (below a few $\mathrm{GeV}/c$), multiple soft collisions of the charm quark with the constituents of the expanding medium lead to energy--momentum exchange with the medium, participation in the collective flow, (partial) charm-quark thermalization, and a significant D-meson \vtwo at low \pt. The charm-hadron \vtwo at intermediate \pt (up to approximately $10~\mathrm{GeV}/c$) is sensitive to the underlying hadronization mechanism. In the presence of a deconfined medium, charm quarks can hadronize through two competing mechanisms: fragmentation (expected to dominate at high \pt), in which a fraction of the quark momentum is transferred to the newly-produced lower-\pt hadron, or recombination (dominant at low and intermediate \pt), in which the quark coalesces with nearby comoving quarks from the medium~\cite{Greco:2003xt,Fries:2003vb,Greco:2003vf} to produce a higher-momentum hadron.  At high \pt (above $10~\mathrm{GeV}/c$), where the effect of radiative energy loss is expected to dominate, measurements of charm-hadron \vtwo can provide insight into the path-length dependence of charm-quark in-medium energy loss~\cite{Gyulassy:2000gk,Shuryak:2001me}.

The theoretical description of heavy-flavor propagation through the QGP is commonly implemented in transport models~\cite{Alberico:2013bza,He:2014cla,Nahrgang:2013xaa,Song:2015sfa,Scardina:2017ipo,Ke:2018tsh} that account for the heavy-quark initial production, diffusion in the medium, hadronization, and subsequent hadronic interactions. In this context, the spatial diffusion coefficient of heavy quarks in the QGP is particularly relevant, since it is related to the relaxation (equilibration) time of heavy quarks in the medium~\cite{Moore:2004tg}. Comparisons of experimental measurements with theoretical predictions — where elastic interactions and hadronization via coalescence are essential to reproduce the observed elliptic flow — and with Bayesian analyses indicate that charm quarks reach (partial) thermalization and that the relaxation time of low-\pt charm quarks is comparable to the QGP lifetime~\cite{ATLAS:2021xtw, ALICE:2021rxa, Xu:2017obm, Ke:2018tsh}.

Measurements of charm-hadron \vn down to \pt close to zero probe the interactions of charm quarks with the medium and provide constraints on the charm-quark relaxation time and the transport properties of the QGP~\cite{Batsouli:2002qf}. Additionally, measurements of \vtwo of \Ds mesons, which contain a strange quark, allow testing the mass and quark-composition dependence of charm-meson \vtwo through comparison with lighter \Dzero and \Dplus mesons. They can shed light on possible differences in the hadronization time of strange and non-strange D mesons within sequential coalescence models, or in their interactions in the hadron-gas phase. In sequential coalescence models, \Ds mesons are expected to hadronize earlier than non-strange D mesons~\cite{10.1007/978-3-030-53448-6_42}, which may result in a reduced buildup of elliptic flow. Moreover, an earlier decoupling of strange mesons from the hadron gas would reduce the flow acquired by \Ds mesons during the late hadronic stage~\cite{He:2012df}. Extending these studies to charm baryons, such as the \Lc, offers a unique opportunity to further constrain the diffusion properties of the QGP and to study the hadronization process of charm quarks. In the light-flavor sector, a characteristic grouping of baryon and meson \vtwo values is observed at intermediate \pt, approximately starting from 3 $\mathrm{GeV}/c$~\cite{STAR:2003wqp,STAR:2005npq,PHENIX:2006dpn,ALICE:2014wao,ALICE:2018yph}. This feature is commonly interpreted as evidence that collective flow develops at the partonic level, and that hadron formation in this momentum region is dominated by quark recombination. Within this hadronization mechanism, the \vtwo of mesons (baryons) reflects the combined \vtwo of their two (three) constituent quarks, resulting in larger \vtwo values for baryons than for mesons and a shift of the \vtwo peak towards higher transverse momenta~\cite{Molnar:2003ff,Fries:2003vb,Voloshin:2002wa}.

In this Letter, we report the measurements of the elliptic flow coefficient of prompt \Dzero, \Dplus, and \Ds mesons and, for the first time, of prompt \Lc baryons in \PbPb collisions at center-of-mass energy per nucleon pair of $\sNN=5.36$ \TeV for the 30--50\% centrality class at midrapidity ($\yrange 0.8$), using the ALICE detector.

The ALICE apparatus, upgraded during the LHC Long Shutdown 2 and operated during LHC Run 3, is described in detail in Ref.~\cite{ALICE:2023udb}. The detectors used in this study are the Inner Tracking System (ITS2)~\cite{ALICE:2013nwm}, a silicon pixel detector based on monolithic active pixel sensors, providing precise track and vertex reconstruction; the Time Projection Chamber (TPC)~\cite{ALICETPC:2020ann}, which is used for tracking and particle identification (PID) via specific energy-loss measurements; and the Time-Of-Flight (TOF) detector~\cite{ALICE:2023udb}, which complements the TPC PID information by measuring the time of flight of charged particles. These detectors are positioned in the central-barrel region, covering the pseudorapidity range $|\eta| < 0.9$, within a solenoidal magnetic field of 0.5~T along the beam axis. The Fast Interaction Trigger (FIT)~\cite{Trzaska:2017reu} system, which includes the FV0 ($2.2 < \eta < 5.1$) and FT0, composed of FT0A ($3.5 < \eta < 4.9$) and FT0C ($-3.3 < \eta < -2.1$), provides fast event selection based on forward multiplicity and collision centrality. The FV0 and FT0 detectors are used to determine the flow vectors, which are essential to measure the \vtwo of charm hadrons.

The measurements are performed on the \PbPb data sample collected in 2023. Events are required to have coincident signals in the FT0A and FT0C, and a reconstructed primary vertex within $\pm$10~cm from the nominal interaction point along the beam axis.  The collision centrality, estimated using the FT0C signal amplitude and expressed as percentiles of the total hadronic cross section, is restricted to the 30--50\% class. A total number of $1.7\times 10^9$ events are selected, corresponding to an integrated luminosity of \mbox{$\mathcal{L}_\mathrm{int} \simeq$ 1.1~nb$^{-1}$}, almost twenty times higher than the previous Run 2 data sample in the same centrality interval. This increase is also made possible by the upgraded readout system, which allows continuous data taking at interaction rates up to 50~kHz in \PbPb collisions.

\begin{sloppypar}
    The \Dzero, \Dplus, \Ds mesons and \Lc baryons, together with their antiparticles, are reconstructed at midrapidity ($\yrange0.8$) in their hadronic decay channels: $\DzerotoKpi$ (branching ratio BR $= (3.95\pm0.03)\%$), $\DplustoKpipi$ (BR $= (9.38\pm0.16)\%$), $\DstophipitoKKpi$ (BR $= (2.22\pm0.06)\%$), and \mbox{$\LctopKpi$} (BR $= (6.28\pm0.32)\%$)~\cite{ParticleDataGroup:2024cfk}. Charm-hadron candidates are built by reconstructing their decay-vertex topology, combining reconstructed charged tracks with the appropriate charge sign. Tracks are required to have \pt $> 0.4~\GeVc$ and $|\eta| < 0.8$, a minimum number of clusters in the ITS2 and TPC detectors, and compatibility within 3 to 5$\sigma$ between the measured and expected signals in both the TPC and TOF detectors (when available) under the relevant PID hypothesis.
\end{sloppypar}

Charm-hadron candidates are selected using multi-class classification models based on machine-learning algorithms, specifically Boosted Decision Trees (BDTs) from the XGBoost package~\cite{10.1145/2939672.2939785}. This approach allows simultaneous suppression of the large combinatorial background and separation of prompt charm hadrons, produced in charm-quark hadronization and excited charm-hadron decays, from non-prompt ones, originating from beauty-hadron decays. BDTs are trained on background, prompt, and non-prompt candidates. Background candidates are selected from regions of the invariant-mass distribution away from the signal peak in data, providing a realistic representation of the background. Prompt and non-prompt candidates are taken from Monte Carlo (MC) simulations of \PbPb collisions generated with Angantyr~\cite{Bierlich:2018xfw} and enriched with heavy-quark pairs from PYTHIA~8.3~\cite{Bierlich:2022pfr}. Kinematic, topological, and PID variables are exploited in the training. Independent BDT models are trained for each charm-hadron species and \pt interval of the measurement.

The \vtwo of the selected charm hadrons is measured as a function of \pt using the scalar-product (SP) method~\cite{Poskanzer:1998yz, Luzum:2012da}:
\begin{equation}
    v_{2} \{\mathrm{SP}\} =
    \frac{
    \langle\!\langle \mathbf{u}_{2} \cdot \mathbf{Q}_{2}^{*\,\mathrm{FT0C}} \rangle\!\rangle
    }{
    R_{2}
    },
    \quad
    R_{2} =
    \sqrt{
    \frac{
    \langle \mathbf{Q}_{2}^{\mathrm{FT0C}} \!\cdot\! \mathbf{Q}_{2}^{*\,\mathrm{FV0A}} \rangle \;
    \langle \mathbf{Q}_{2}^{\mathrm{FT0C}} \!\cdot\! \mathbf{Q}_{2}^{*\,\mathrm{TPC}} \rangle
    }{
    \langle \mathbf{Q}_{2}^{\mathrm{FV0A}} \!\cdot\! \mathbf{Q}_{2}^{*\,\mathrm{TPC}} \rangle
    }
    }\;,
    \label{eq:vtwo}
\end{equation}

where $\mathbf{u}_{2}$ is the second-harmonic azimuthal unit vector of the reconstructed charm-hadron candidate and $\mathbf{Q}_{2}^{\mathrm{Det}}$ is the second-harmonic flow vector measured with the detector
``Det'', scaled by the charged-particle multiplicity for the TPC or by the total signal amplitude for the FT0C and FV0A. The $\mathbf{Q}_{2}^{\mathrm{Det}}$ vectors are calibrated to correct for non-uniform-acceptance effects~\cite{Selyuzhenkov:2007zi}. Single angular brackets denote an average over all events, while double angular brackets denote an average over all events and candidates in that \pt interval. The resolution term, $R_{2}$, at the denominator of Eq.~\ref{eq:vtwo} is computed using the FT0C, FV0A, and TPC detectors. This configuration ensures a pseudorapidity gap of at least 1.3 between the region where charm hadrons are reconstructed and the FT0C used for flow-vector determination, thereby suppressing spurious contributions (non-flow).

Since the \vtwo of charm hadrons cannot be extracted on a candidate-by-candidate basis, the inclusive \vtwo signal, containing both prompt and non-prompt candidates, is obtained from a simultaneous fit to the invariant-mass distribution and the \vtwo of the selected candidates as a function of the invariant mass, separately for each \pt interval~\cite{ALICE:2023gjj, ALICE:2020iug}. 
To obtain the prompt charm-hadron \vtwo, the inclusive \vtwo signal is corrected for the residual contribution from beauty-hadron decays. The fraction of non-prompt candidates, \fnonprompt, is determined in each \pt interval using a data-driven approach~\cite{cutvariationdmesprod}. In this procedure, raw yields extracted from invariant-mass fits and acceptance-times-efficiency factors for both prompt and non-prompt sources, obtained from MC simulations, are evaluated for progressively tighter BDT selections. The variation of the raw yields and acceptance-times-efficiency factors with the selection criteria allows estimating the prompt and non-prompt fractions for a given BDT selection. The prompt charm-hadron \vtwo is finally extrapolated from a linear fit to the inclusive \vtwo as a function of the non-prompt fraction obtained for sets of non-overlapping BDT selections, evaluating the prompt charm-hadron \vtwo at a null non-prompt fraction.

The main sources of systematic uncertainties affecting the \vtwo measurements are associated with: i) the simultaneous-fit procedure, ii) the estimate of \fnonprompt, and iii) the centrality dependence of $R_{2}$. The systematic uncertainty associated with the fit is assessed by repeating the  procedure with different configurations of the line shapes used to describe the signal and background components, binning, and fit range. A systematic uncertainty of $0.002$–$0.030$ depending on the momentum and particle species is assigned to the measured \vtwo based on the RMS and the shift in trials relative to the default case. Similarly, the systematic uncertainty on the estimate of \fnonprompt is assigned by varying the BDT selections included in the data-driven approach described above, resulting in a systematic uncertainty on the measured \vtwo of $0.002$–$0.005$. A relative systematic uncertainty of 0.2\% is assigned to the \vtwo by estimating the difference between the centrality-integrated $R_{2}$ and the value obtained when weighting it by the charm-hadron yields in narrow centrality intervals. More details on the analysis procedure are provided in the \mbox{\hyperref[endmatter]{End Matter}}.

\begin{figure}[htb!]
    \centering
    \includegraphics[width=0.62\textwidth]{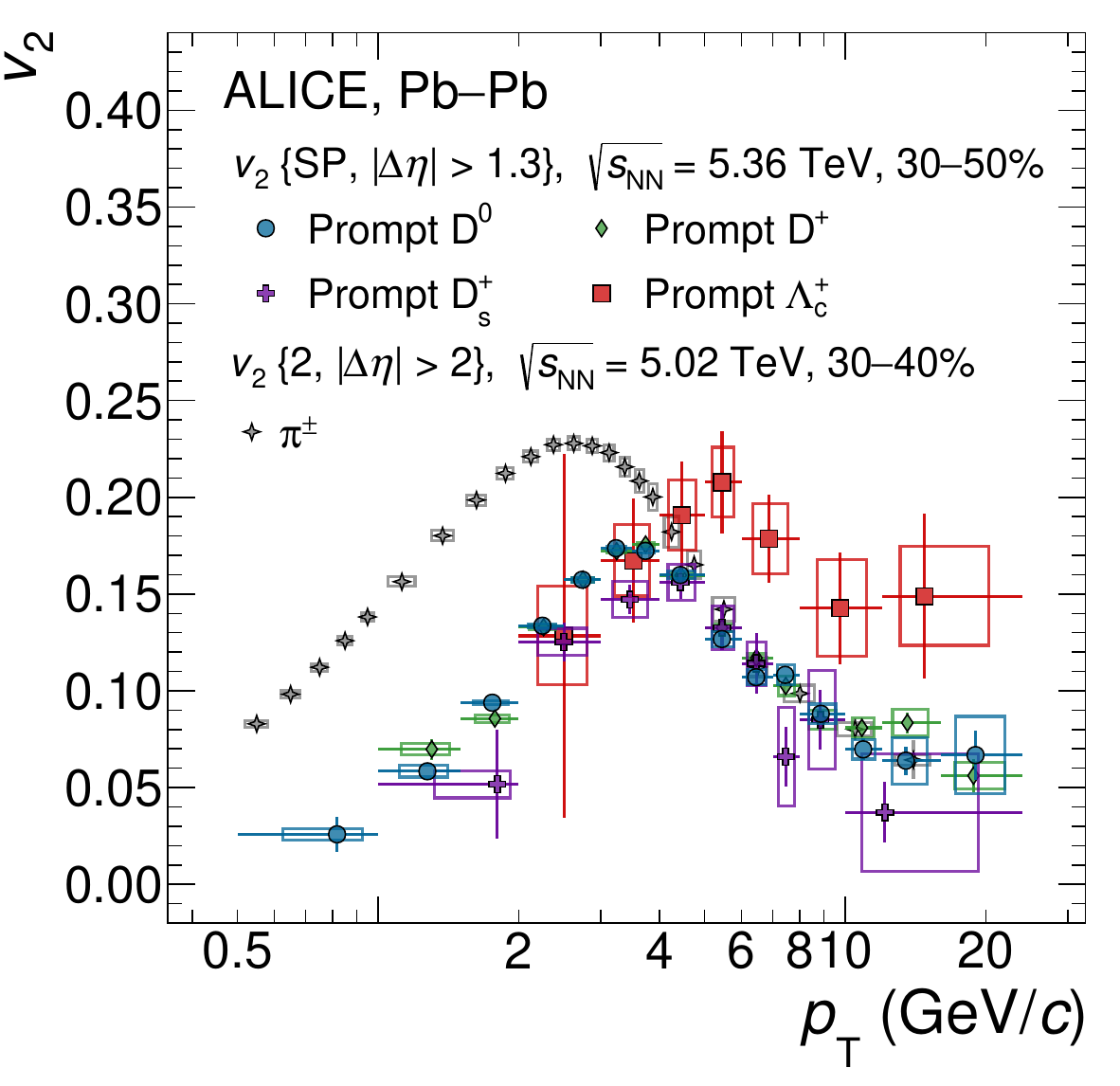}
    \caption{Elliptic flow of prompt \Dzero, \Dplus, \Ds mesons and \Lc baryons at midrapidity (\yrange0.8) in \PbPb collisions at $\sNN = 5.36$~TeV in the 30$-$50\% centrality class as a function of the charm-hadron \pt. The \vtwo of charged pions measured at $\sNN = 5.02$~TeV in the 30$-$40\% centrality class~\cite{ALICE:2018yph} is also shown for comparison.}
    \label{fig:v2_Dmesons_Lc}
\end{figure}

The elliptic flow of prompt \Dzero, \Dplus, \Ds mesons and \Lc baryons at midrapidity (\yrange 0.8) in \PbPb collisions at $\sNN = 5.36$~TeV in the 30--50\% centrality class as a function of \pt is shown in Fig.~\ref{fig:v2_Dmesons_Lc}. The statistical and systematic uncertainties are represented as error bars and empty boxes, respectively. In the same figure, the \vtwo of charged pions (\yrange 0.5) measured at $\sNN = 5.02$~TeV in the 30--40\% centrality class~\cite{ALICE:2018yph} is shown. A positive \vtwo is measured for all charm-hadron species over all \pt, and the \vtwo for \Dzero mesons is measured for \pt lower than $1~\GeVc$ for the first time.

At low transverse momentum ($\pt < 4~\GeVc$), the charm-hadron \vtwo increases with \pt and exhibits a clear mass ordering ($\vtwo(\pipm) > \vtwo(\rm D)$). This behavior, already observed in heavy-ion collisions for light- and heavy-flavor hadrons~\cite{ALICE:2018yph, ALICE:2020chv, ALICE:2018gif, STAR:2004jwm}, highlights the interplay between radial and elliptic flow inside a hydrodynamic expanding medium. The prompt \Dzero- and \Dplus-meson \vtwo are compatible within the experimental uncertainties.  The measured $\vtwo$ values indicate a lower elliptic flow of \Ds mesons compared to \Dzero mesons in the $1 < \pt < 5~\GeVc$ interval. The probability of observing the measured deviation obtained with a one-sided log-likelihood ratio test is $2.6\sigma$, assuming the null hypothesis that the difference between \Dzero and \Ds \vtwo is on average zero in the considered \pt interval. In the significance computation, the sources of statistical and systematic uncertainty on the simultaneous fit and the estimate of \fnonprompt are treated as fully uncorrelated as a function of \pt and between \Dzero and \Ds. The uncertainty associated with the resolution factor $R_{2}$ is considered fully correlated. According to Ref.~\cite{He:2012df}, a difference could arise from an early kinetic freeze-out of strange relative to non-strange hadrons during the hadronic phase of the collision: this would reduce the amount of flow transferred to the \Ds through multiple low-energy interactions with the hadronic gas compared to the \Dzero and \Dplus mesons. However, recent femtoscopy measurements of D-meson scattering length indicate a negligible strong interaction between D mesons and pions~\cite{ALICE:2024bhk}. An alternative interpretation could be related to a sequential hadronization scenario~\cite{10.1007/978-3-030-53448-6_42}, in which strange D mesons are expected to coalesce on a shorter timescale than non-strange ones, resulting in a smaller degree of charm-quark thermalization and a smaller \Ds \vtwo compared to \Dzero. The present experimental uncertainties, however, do not allow for a definitive conclusion.
For $\pt > 4$~\GeVc, the \vtwo of D mesons smoothly decreases with increasing \pt, following the trend and magnitude of \pipm \vtwo. In the $4 < \pt < 12~\GeVc$ interval, the \vtwo of \Lc baryons is larger than that of \Dzero mesons with a $3.7\sigma$ significance, representing the first evidence for baryon--meson splitting in the heavy-flavor sector. The origin of the \vtwo splitting between baryons and mesons, already observed for lighter particles~\cite{ALICE:2018yph}, can be attributed to the formation of hadrons via recombination, which is expected to be a significant process in the presence of a dense color-deconfined medium. Low-\pt charm quarks are expected to predominantly hadronize through recombination with light quarks from the medium that are close in phase space, forming higher-momentum hadrons. This mechanism results in the observed \mbox{baryon--meson} splitting of \vtwo at intermediate \pt, with baryons exhibiting a larger \vtwo and a peak shifted to higher \pt compared to mesons, reflecting their different number of constituent quarks. This result provides strong evidence of the partonic origin of charm-hadron elliptic flow. For \pt higher than 10~\GeVc, a similar non-zero \vtwo is observed for all charm-meson species, which can be attributed to a common energy loss experienced by high-momentum partons. In this \pt region, a hint of a larger \Lc \vtwo compared to that of D mesons is found, despite the expected similar sensitivity of baryons and mesons to path-length–dependent energy loss~\cite{ALICE:2021bib}. Given the current experimental uncertainties, no definitive conclusion can be drawn.

\begin{figure}[htbp]
    \centering
    \includegraphics[width=0.6\textwidth]{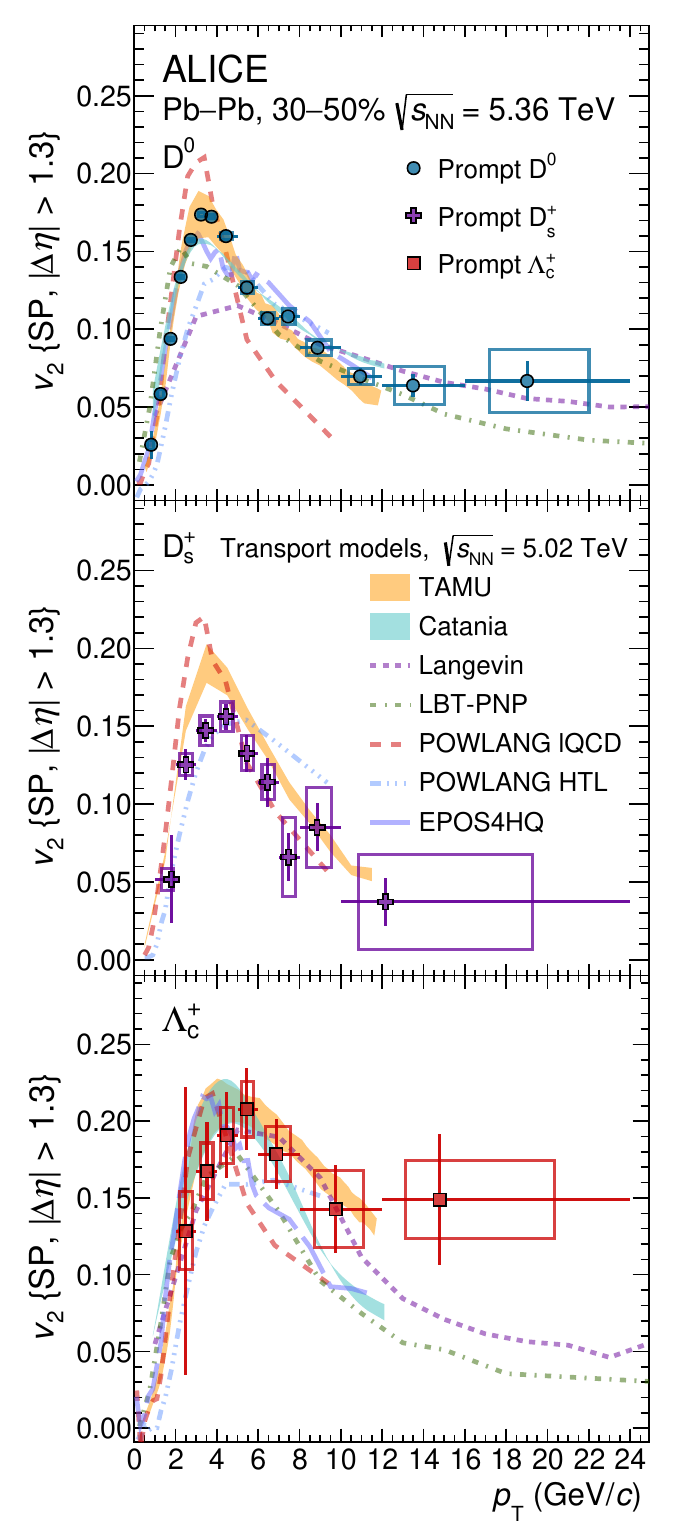}
    \caption{Elliptic flow of prompt \Dzero (top), \Ds (middle), and \Lc (bottom) hadrons at midrapidity (\yrange 0.8) in \PbPb collisions at $\sNN = 5.36$~TeV in the 30–-50\% centrality class as a function of \pt. The data are compared with predictions from TAMU~\cite{He:2019vgs}, Catania~\cite{Scardina:2017ipo,Sambataro:2024mkr}, POWLANG~\cite{Beraudo:2022dpz,Beraudo:2023nlq}, LBT-PNP~\cite{Ke:2018tsh,Xing:2021xwc}, EPOS4HQ~\cite{Zhao:2024ecc}, and Langevin~\cite{Li:2021xbd,Li:2020kax} transport models at $\sNN = 5.02$~TeV.}
    \label{fig:v2_D0_Lc_compare_models}
\end{figure}

Figure~\ref{fig:v2_D0_Lc_compare_models} presents a comparison of the measured \vtwo of prompt \Dzero (top panel), \Ds (middle panel), and \Lc (bottom panel) hadrons with several theoretical model calculations. Under the assumption that the heavy-quark masses exceed the magnitude of typical momentum transfers in elastic interactions with the medium, the heavy-quark transport in the QGP is generally treated using either the Boltzmann equation, as in LBT-PNP~\cite{Ke:2018tsh,Xing:2021xwc} and EPOS4HQ~\cite{Zhao:2024ecc}, or the Langevin approach. Several models further extend this description to include the transport of heavy-flavor hadrons in the hadron-gas phase. A comprehensive overview of the computation and treatment of the spatial diffusion coefficient in heavy-quark transport models is provided in Ref.~\cite{Rapp:2018qla}. The interactions between the charm quarks and the medium constituents are described either as the result of multiple elastic collisions, as implemented in the TAMU~\cite{He:2019vgs}, Catania~\cite{Scardina:2017ipo,Sambataro:2024mkr}, and POWLANG~\cite{Beraudo:2022dpz,Beraudo:2023nlq} models, or by including both elastic and radiative processes, as implemented in EPOS4HQ~\cite{Zhao:2024ecc}, Langevin~\cite{Li:2021xbd,Li:2020kax}, and LBT-PNP~\cite{Ke:2018tsh,Xing:2021xwc}. The variants of the POWLANG model differ in the transport coefficients, which are derived either from weak-coupling Hard-Thermal-Loop (HTL) calculations or from lattice-QCD (lQCD) computations. Models also differ in their treatment of initial-state effects such as nuclear parton distribution functions (nPDFs) and shadowing. The Catania model implements the space--time evolution of the medium through Boltzmann quasi-particle transport, while the others consider ideal or viscous hydrodynamics. Finally, all the discussed models implement the hadronization of charm quarks as the interplay of fragmentation and coalescence mechanisms, even though their realizations differ. In particular, instantaneous recombination based on the Wigner function formalism is considered in all models, except for TAMU and POWLANG. In the former, heavy quarks coalesce with thermal light quarks from the medium according to their space--momentum correlation to form unstable resonant states, which eventually decay in the observed ground state. In POWLANG, in-medium hadronization is modeled via local color neutralization, where charm quarks recombine with nearby quarks or diquarks to form color-singlet clusters that subsequently decay into charmed mesons or baryons. Most of the model predictions provide a qualitative description of the measured \vtwo of D mesons and \Lc baryons. TAMU exhibits the best overall agreement with the data, although it slightly overestimates the measured \Ds \vtwo at intermediate \pt.
The Catania model describes the \vtwo of \Dzero over almost the full \pt range of the measurement with the exception of the \vtwo peak, while it over- or underestimates the \vtwo of \Lc at low or high \pt
, respectively. Predictions from POWLANG (HTL and lQCD) and LBT-PNP qualitatively reproduce the trends observed in data but tend to overshoot or undershoot the measurements depending on the \pt and the particle species. Finally, EPOS4HQ and Langevin qualitatively describe either \Dzero or \Lc, but tend to underestimate the other hadron. A summary of the data–model compatibility, expressed in terms of $\chi^{2}/\mathrm{ndf}$ values computed following the procedure detailed in previous publications~\cite{ALICE:2021rxa}, is reported in Table~\ref{tab:chi2_models}. Given the different \pt coverage of the various model predictions, a common range of $0.5<\pt<10$~\GeVc, $1<\pt<10$~\GeVc, and $2 < \pt <8$~\GeVc is considered for \Dzero, \Ds, and \Lc respectively, when calculating the $\chi^{2}/\mathrm{ndf}$. Theoretical uncertainties are included in the $\chi^{2}/\mathrm{ndf}$ calculations, when available~\cite{ALICE:2021rxa}.

\begin{table}
    \caption{Summary of the $\chi^{2}/\mathrm{ndf}$  obtained for different model predictions in the $0.5<\pt<10$~\GeVc  and $2 < \pt <8$~\GeVc intervals for \vtwo of  prompt D mesons and \Lc baryons, respectively. Theoretical uncertainties are included in the $\chi^{2}/\mathrm{ndf}$ calculations, when available~\cite{ALICE:2021rxa}.}
    \centering
    \begin{tabular}{l | c c c}
        \toprule
        Transport model                                     & $\chi^{2}/\mathrm{ndf}$ $(\Dzero)$ & $\chi^{2}/\mathrm{ndf}$ $ (\Ds)$ & $\chi^{2}/\mathrm{ndf}$ $ (\Lc)$ \\
        \midrule
        Catania~\cite{Scardina:2017ipo,Sambataro:2024mkr}   & 13.81                              & --                               & 0.24                             \\
        EPOS4HQ~\cite{Zhao:2024ecc}                         & 31.74                              & --                               & 0.96                             \\
        Langevin~\cite{Li:2021xbd,Li:2020kax}               & 298.11                             & --                               & 0.08                             \\
        LBT-PNP~\cite{Ke:2018tsh,Xing:2021xwc}              & 117.25                             & --                               & 0.59                             \\
        POWLANG-HTL~\cite{Beraudo:2022dpz,Beraudo:2023nlq}  & 255.30                             & 2.11                             & 0.90                             \\
        POWLANG-lQCD~\cite{Beraudo:2022dpz,Beraudo:2023nlq} & 151.83                             & 7.94                             & 2.07                             \\
        TAMU~\cite{He:2019vgs}                              & 0.57                               & 2.16                             & 0.26                             \\
        \bottomrule
    \end{tabular}
    \label{tab:chi2_models}
\end{table}

In this Letter, we present the measurements of the elliptic flow of prompt \Dzero, \Dplus, \Ds mesons and, for the first time, prompt \Lc baryons in semi-central \PbPb collisions at $\sNN=5.36$ \TeV at midrapidity (\yrange 0.8). The \vtwo values of \Dzero and \Dplus mesons are compatible within uncertainties in the full \pt interval of the measurement, while a hint of deviation of $2.6\sigma$ in $1 <\pt < 5$~\GeVc is observed between \Dzero and \Ds mesons. The prompt \Lc-baryon \vtwo exceeds that of D mesons for $4 < \pt < 12$~\GeVc with a significance of 3.7$\sigma$. The baryon--meson splitting in the charm sector indicates a partonic origin of the flow. The measurements are compared with several transport model predictions. The models generally capture the qualitative features of the data. Among them, TAMU provides the best quantitative description across the different charm-hadron species. 


\newenvironment{acknowledgement}{\relax}{\relax}
\begin{acknowledgement}
    \section*{Acknowledgements}

The ALICE Collaboration would like to thank all its engineers and technicians for their invaluable contributions to the construction of the experiment and the CERN accelerator teams for the outstanding performance of the LHC complex.
The ALICE Collaboration gratefully acknowledges the resources and support provided by all Grid centres and the Worldwide LHC Computing Grid (WLCG) collaboration.
The ALICE Collaboration acknowledges the following funding agencies for their support in building and running the ALICE detector:
A. I. Alikhanyan National Science Laboratory (Yerevan Physics Institute) Foundation (ANSL), State Committee of Science and World Federation of Scientists (WFS), Armenia;
Austrian Academy of Sciences, Austrian Science Fund (FWF): [M 2467-N36] and Nationalstiftung f\"{u}r Forschung, Technologie und Entwicklung, Austria;
Ministry of Communications and High Technologies, National Nuclear Research Center, Azerbaijan;
Rede Nacional de Física de Altas Energias (Renafae), Financiadora de Estudos e Projetos (Finep), Funda\c{c}\~{a}o de Amparo \`{a} Pesquisa do Estado de S\~{a}o Paulo (FAPESP) and The Sao Paulo Research Foundation  (FAPESP), Brazil;
Bulgarian Ministry of Education and Science, within the National Roadmap for Research Infrastructures 2020-2027 (object CERN), Bulgaria;
Ministry of Education of China (MOEC) , Ministry of Science \& Technology of China (MSTC) and National Natural Science Foundation of China (NSFC), China;
Ministry of Science and Education and Croatian Science Foundation, Croatia;
Centro de Aplicaciones Tecnol\'{o}gicas y Desarrollo Nuclear (CEADEN), Cubaenerg\'{\i}a, Cuba;
Ministry of Education, Youth and Sports of the Czech Republic, Czech Republic;
The Danish Council for Independent Research | Natural Sciences, the VILLUM FONDEN and Danish National Research Foundation (DNRF), Denmark;
Helsinki Institute of Physics (HIP), Finland;
Commissariat \`{a} l'Energie Atomique (CEA) and Institut National de Physique Nucl\'{e}aire et de Physique des Particules (IN2P3) and Centre National de la Recherche Scientifique (CNRS), France;
Bundesministerium f\"{u}r Forschung, Technologie und Raumfahrt (BMFTR) and GSI Helmholtzzentrum f\"{u}r Schwerionenforschung GmbH, Germany;
National Research, Development and Innovation Office, Hungary;
Department of Atomic Energy Government of India (DAE), Department of Science and Technology, Government of India (DST), University Grants Commission, Government of India (UGC) and Council of Scientific and Industrial Research (CSIR), India;
National Research and Innovation Agency - BRIN, Indonesia;
Istituto Nazionale di Fisica Nucleare (INFN), Italy;
Japanese Ministry of Education, Culture, Sports, Science and Technology (MEXT) and Japan Society for the Promotion of Science (JSPS) KAKENHI, Japan;
Consejo Nacional de Ciencia (CONACYT) y Tecnolog\'{i}a, through Fondo de Cooperaci\'{o}n Internacional en Ciencia y Tecnolog\'{i}a (FONCICYT) and Direcci\'{o}n General de Asuntos del Personal Academico (DGAPA), Mexico;
Nederlandse Organisatie voor Wetenschappelijk Onderzoek (NWO), Netherlands;
The Research Council of Norway, Norway;
Pontificia Universidad Cat\'{o}lica del Per\'{u}, Peru;
Ministry of Science and Higher Education, National Science Centre and WUT ID-UB, Poland;
Korea Institute of Science and Technology Information and National Research Foundation of Korea (NRF), Republic of Korea;
Ministry of Education and Scientific Research, Institute of Atomic Physics, Ministry of Research and Innovation and Institute of Atomic Physics and Universitatea Nationala de Stiinta si Tehnologie Politehnica Bucuresti, Romania;
Ministerstvo skolstva, vyskumu, vyvoja a mladeze SR, Slovakia;
National Research Foundation of South Africa, South Africa;
Swedish Research Council (VR) and Knut \& Alice Wallenberg Foundation (KAW), Sweden;
European Organization for Nuclear Research, Switzerland;
Suranaree University of Technology (SUT), National Science and Technology Development Agency (NSTDA) and National Science, Research and Innovation Fund (NSRF via PMU-B B05F650021), Thailand;
Turkish Energy, Nuclear and Mineral Research Agency (TENMAK), Turkey;
National Academy of  Sciences of Ukraine, Ukraine;
Science and Technology Facilities Council (STFC), United Kingdom;
National Science Foundation of the United States of America (NSF) and United States Department of Energy, Office of Nuclear Physics (DOE NP), United States of America.
In addition, individual groups or members have received support from:
FORTE project, reg.\ no.\ CZ.02.01.01/00/22\_008/0004632, Czech Republic, co-funded by the European Union, Czech Republic;
European Research Council (grant no. 950692), European Union;
Deutsche Forschungs Gemeinschaft (DFG, German Research Foundation) ``Neutrinos and Dark Matter in Astro- and Particle Physics'' (grant no. SFB 1258), Germany;
FAIR - Future Artificial Intelligence Research, funded by the NextGenerationEU program (Italy).

\end{acknowledgement}

\bibliographystyle{utphys}   
\bibliography{bibliography}

\newpage
\appendix
\makeatletter
\renewcommand{\thesubsection}{\arabic{subsection}}
\makeatother

\section*{End Matter}
\label{endmatter}
\renewcommand{\thefigure}{\arabic{figure}}
\renewcommand{\theequation}{\arabic{equation}}
\setcounter{figure}{0}
\setcounter{equation}{0}

\subsection*{Track quality criteria}
\label{appendixrec}
All charm-hadron candidates are built by combining pairs or triplets of tracks with appropriate charge sign, which satisfy quality criteria based on their kinematic, topological, and PID information. These selections are intended to suppress the background arising from erroneous combinations of tracks. The tracks are required to have $|\eta|<0.8$ and $\pt>0.4~\GeVc$, at least 70 TPC space points (out of 159), and at least four ITS hits, one of which is in the three innermost layers. The distance of closest approach to the primary vertex must be less than 15 or 25~$\upmu$m, depending on the hadron species. PID selections require compatibility within $3$–$5\sigma$ between the measured and expected TPC and TOF signals.

\subsection*{Analysis strategy}
\label{appendixanalysis}
This section complements the description of the analysis procedure by outlining the fundamental steps of the analysis chain: i) the training of the machine-learning algorithm, ii) the signal \vtwo extraction via simultaneous fits, iii) the estimate of the prompt fraction based on a data-driven method, and iv) the prompt charm-hadron \vtwo extrapolation as a function of the prompt fraction.

The multi-class BDT models are trained using kinematic, topological, and PID information of prompt and non-prompt candidates, and background candidates. The signal candidates are taken from MC simulations of \PbPb\ collisions generated with Angantyr~\cite{Bierlich:2018xfw}, enriched with PYTHIA~8.3~\cite{Bierlich:2022pfr} events containing a heavy-quark pair, employing the color reconnection beyond-leading-color approximation (Mode~2)~\cite{Christiansen:2015yqa}. The particle transport and the detector response are simulated using the GEANT~4 transport package~\cite{GEANT4:2002zbu}. The background sample of candidates is selected from the invariant mass sidebands more than $5\sigma$ away from the signal peak. The most discriminating features exploited by the algorithm are related to the displacement of the charm-hadron decay (secondary) vertex from the primary vertex of the collision. These include the impact parameters of the tracks composing the candidate, the distance between the secondary and primary vertices, and the cosine of the pointing angle between the vector connecting the two vertices and the reconstructed momentum vector of the candidate. In addition, TPC and TOF PID information of the tracks is included to improve the separation between signal and background candidates. When applied, the BDT assigns to each candidate a score representing its probability of belonging to each class (prompt signal, non-prompt signal, and background), which is later used to select charm-hadron candidates.

\begin{figure}[htb!]
    \centering
    \includegraphics[width=0.65\textwidth]{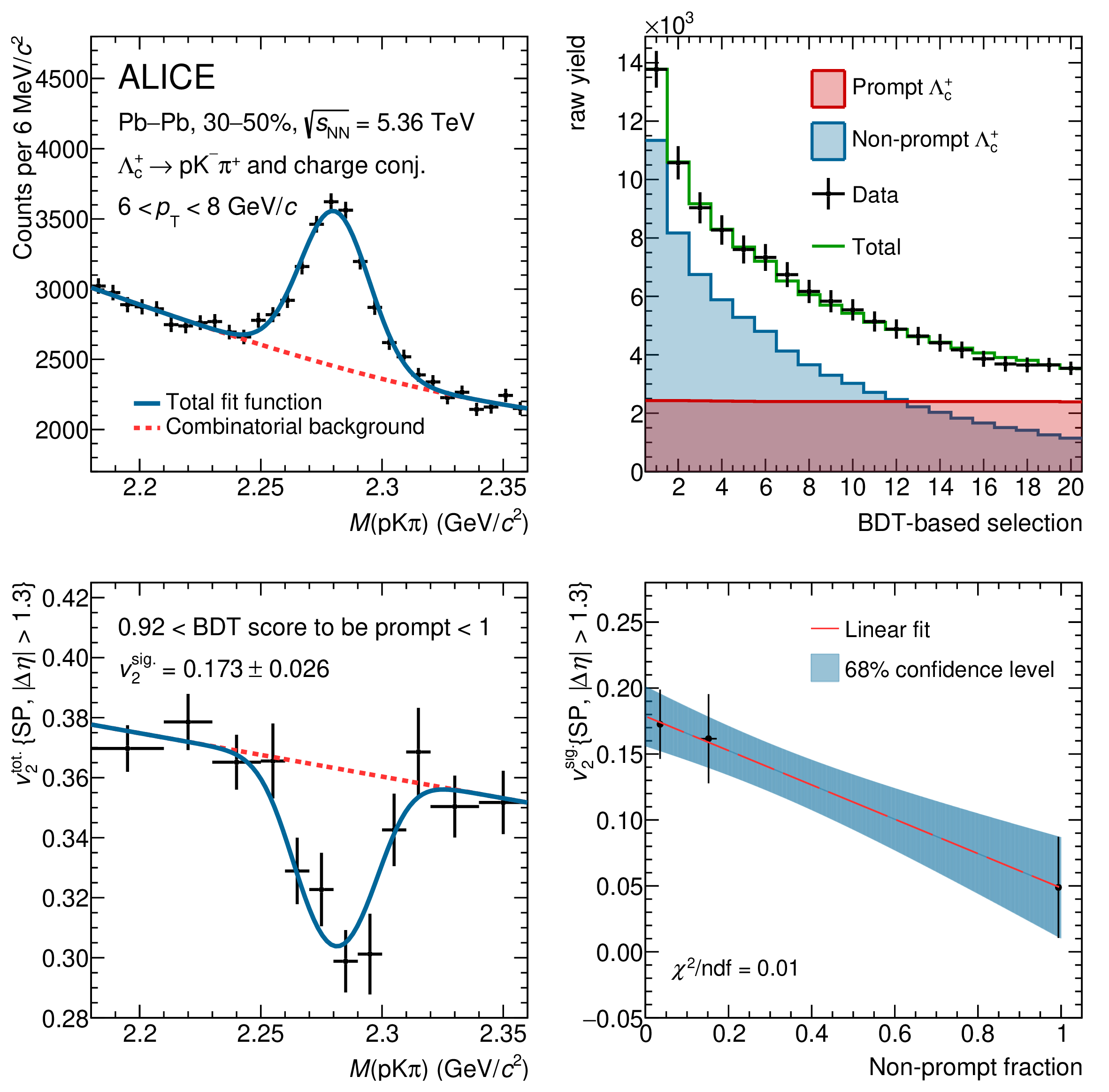}
    \caption{Left: simultaneous fit of the candidate yield (top) and total $\vtwo$ as a function of the candidate invariant mass (bottom) of \Lc baryons in $6 < \pt < 8~\GeVc$ in \PbPb collisions for the 30--50\% centrality class. The combinatorial background and the total fit functions are depicted as dotted red curves and blue lines, respectively. The BDT-based selection and the corresponding signal \vtwo are reported in the bottom panel. Top right: raw-yield distribution as a function of the minimum threshold on the BDT prompt score of \Lc baryons in $6 < \pt < 8~\GeVc$ for the 30--50\% centrality class. The prompt and non-prompt populations, along with their sum, are reported in red, blue, and green, respectively. Bottom right: linear fit to the signal $\vtwo$ of \Lc baryons in $6 < \pt < 8~\GeVc$ for the 30--50\% centrality class as a function of the non-prompt fraction. The blue band represents the $1\sigma$ confidence interval obtained from the fit.}
    \label{fig:performance}
\end{figure}

The top- and bottom-left panels of Fig~\ref{fig:performance} show an example of the simultaneous fit of the reconstructed candidate yield and the total \vtwo ($v_{2}^{\mathrm{tot}}$) as a function of the candidate invariant mass for the \Lc baryons in the $4<\pt<5~\GeVc$ interval in the 30--50\% centrality class. The total \vtwo decreases significantly in the region of the $\Lc$ peak in the invariant-mass distribution, indicating a distinct elliptic flow for the signal compared to the combinatorial background. The functional forms used to describe the invariant mass and the \vtwo distributions are composed of a Gaussian term to describe the signal peak and an exponential to describe the background. Similarly, the depletion in the distribution of \vtwo as a function of the invariant mass is modeled with a Gaussian term and a linear function to describe the background. The \vtwo coefficient extracted from the fit is expressed as the sum of the \vtwo of the charm-hadron signal ($v_{2}^{\text{sig}}$) and that of background ($v_{2}^{\text{bkg}}$) weighted by the signal ($N_{\text{sig}}$) and background ($N_{\text{bkg}}$) yields, respectively:
\begin{equation}
    \label{eq:fitvnvsmass}
    v_{2}^{\mathrm{tot}} = v_{2}^{\text{sig}} \frac{N_{\text{sig}}}{N_{\text{sig}} + N_{\text{bkg}}}
    + v_{2}^{\text{bkg}} \frac{N_{\text{bkg}}}{N_{\text{sig}} + N_{\text{bkg}}}~.
\end{equation}

Once extracted from the simultaneous fit procedure, the charm-hadron \vtwo has to be corrected by the residual contamination arising from the contribution of non-prompt charm hadrons. The fraction of non-prompt candidates (\fnonprompt) in the selected sample of charm hadrons is estimated via a data-driven approach~\cite{npd0v2}. In this method, a series of data sub-samples with different abundances of prompt and non-prompt candidates is built by progressively tightening the selection on the BDT score associated with the probability of being a prompt or non-prompt candidate. For each subsample i, the raw yields $\rawY{i}$ and the acceptance-times-efficiency factors of prompt ($\effP{i}$) and non-prompt ($\effNP{\rm i}$) charm hadrons are computed and can be related to the corrected yield of prompt ($N_\mathrm{prompt}$) and non-prompt ($N_{\rm non\text{-}prompt}$) charm hadrons:
\begin{equation}
    \effP{i}\times N_\mathrm{prompt} +  \effNP{\rm i}\times N_\mathrm{non\text{-}prompt} = \rawY{i}\;,
    \label{eq:eq_set}
\end{equation}

enabling the construction of a system of equations. Given the experimental uncertainty associated with $\rawY{i}$ and the $(\AccEff)_{\mathrm{i}}$ factors, the system is approximately solved as a function of $N_\mathrm{prompt}$ and $N_\mathrm{non\text{-}prompt}$ to compute \fnonprompt for a given BDT-based selection as:
\begin{equation}
    f_{\rm non-prompt} = \frac{\effNP{\rm i} \times N_\mathrm{non\text{-}prompt}}{\effNP{\rm i} \times N_\mathrm{non\text{-}prompt}+\effP{i} \times N_\mathrm{prompt}}\;.
\end{equation}
In the top-right panel of Fig~\ref{fig:performance}, the raw yields of \Lc baryons are shown as a function of the minimum BDT score threshold of prompt \Lc baryons in $6 < \pt < 8~\GeVc$ for the 30--50\% centrality class. The raw yields, represented by the black markers, decrease smoothly with increasing minimum threshold, corresponding to a decreasing non-prompt contribution, depicted in blue, relative to the prompt one, depicted in red. The sum of prompt and non-prompt corrected yields obtained by solving the system of equations is reported by the green line.

Given the scalar nature of \vtwo, the inclusive signal \vtwo can be interpreted as the linear sum of the \vtwo of prompt and non-prompt charm hadrons. To separate the two contributions, the signal \vtwo is extracted over several non-overlapping intervals of the BDT score to be prompt charm hadrons in order to obtain highly prompt- and non-prompt-enhanced samples. For each BDT selection, the non-prompt fraction is estimated following the aforementioned data-driven procedure. The prompt charm-hadron \vtwo is then extrapolated to $f_{\rm non\text{-}prompt} = 0$ ($f_{\rm prompt} = 1$) using a linear fit. The bottom-right panel of Fig~\ref{fig:performance} shows an example of a linear fit to the signal \vtwo as a function of the non-prompt fraction for the \Lc baryons in $6 < \pt < 8~\GeVc$ for the 30--50\% centrality class. The linear fit is reported as a dotted red line together with a blue band representing the $1\sigma$ confidence interval, which is considered as the statistical uncertainty associated with the prompt charm-hadron \vtwo.

Finally, within each \pt interval, the average transverse momentum of the signal candidates is computed using the sPlot package~\cite{Pivk:2004ty} to statistically separate the signal and background contributions via an unbinned fit to the invariant-mass distribution.
The resulting $\langle \pt \rangle$ value is used as the horizontal coordinate of the data points in Figs.~\ref{fig:v2_Dmesons_Lc} and~\ref{fig:v2_D0_Lc_compare_models}.

\subsection*{Systematic uncertainties}
\label{appendixsyst}
The sources of systematic uncertainty affecting the estimate of the \vtwo of charm hadrons are related to: i) the simultaneous-fit procedure, ii) the estimate of the non-prompt fraction (\fnonprompt), and iii) the centrality dependence of the resolution factor $R_{2}$.

For the first two sources, the assigned systematic uncertainty is defined as the quadratic sum of the RMS of the distribution of trial \vtwo values obtained by appropriately varying different analysis parameters within a given range and the shift of their mean relative to the reference value. Since the three sources are considered fully uncorrelated, the total systematic uncertainty is obtained by summing the individual contributions in quadrature.

The first contribution arises from the choice of the configuration used in the simultaneous fit. Its impact is evaluated by repeating the \vtwo extraction while varying the functional forms (polynomial or exponential) employed to model the combinatorial background and its associated \vtwo, as well as the binning of the invariant-mass spectra and the fit range. The resulting variation is taken as the systematic uncertainty, which ranges from 0.002 to 0.030, depending on the particle species and the \pt interval.

The second source is associated with the determination of the non-prompt fraction, \fnonprompt, which enters the correction procedure used to extract the prompt charm-hadron \vtwo. Its contribution is evaluated by varying the set of BDT selections employed in the data-driven method based on a system of equations for the simultaneous determination of the prompt and non-prompt yields~\cite{npd0v2}. Specifically, subsets of selections are added or removed, and the \fnonprompt estimation is repeated accordingly. The prompt \vtwo extrapolation is then performed again and the resulting variation defines the associated systematic uncertainty, which ranges from 0.002 to 0.005, depending on the particle species and \pt.

The third source originates from the fact that the adopted $R_{2}$ value is computed as the arithmetic average of the resolution factor $R_{2}$ in $1\%$-wide centrality intervals, whereas charm-hadron production exhibits a non-uniform dependence on centrality within a given class. To quantify this effect, raw charm-hadron yields are determined at different \pt for $1\%$-wide centrality bins within the $30$–$50\%$ interval and used as weights to compute a yield-weighted average of $R_{2}$~\cite{ALICE:2021rxa}. The assigned relative systematic uncertainty is 0.2\% and is independent of the particle species and the \pt interval.

\clearpage

\section{The ALICE Collaboration}
\label{app:collab}
\begin{flushleft} 
\small

D.A.H.~Abdallah\,\orcidlink{0000-0003-4768-2718}\,$^{\rm 134}$, 
I.J.~Abualrob\,\orcidlink{0009-0005-3519-5631}\,$^{\rm 112}$, 
S.~Acharya\,\orcidlink{0000-0002-9213-5329}\,$^{\rm 49}$, 
K.~Agarwal\,\orcidlink{0000-0001-5781-3393}\,$^{\rm II,}$$^{\rm 23}$, 
G.~Aglieri Rinella\,\orcidlink{0000-0002-9611-3696}\,$^{\rm 32}$, 
L.~Aglietta\,\orcidlink{0009-0003-0763-6802}\,$^{\rm 24}$, 
N.~Agrawal\,\orcidlink{0000-0003-0348-9836}\,$^{\rm 25}$, 
Z.~Ahammed\,\orcidlink{0000-0001-5241-7412}\,$^{\rm 132}$, 
S.~Ahmad\,\orcidlink{0000-0003-0497-5705}\,$^{\rm 15}$, 
I.~Ahuja\,\orcidlink{0000-0002-4417-1392}\,$^{\rm 36}$
% ...

\section*{Affiliation Notes}

$^{\rm I}$ Deceased\\
$^{\rm II}$ Also at: INFN Trieste\\
$^{\rm III}$ Also at: Fondazione Bruno Kessler (FBK), Trento, Italy\\
$^{\rm IV}$ Also at: Max-Planck-Institut fur Physik, Munich, Germany\\
$^{\rm V}$ Also at: Czech Technical University in Prague (CZ)\\
$^{\rm VI}$ Also at: Instituto de Fisica da Universidade de Sao Paulo\\
$^{\rm VII}$ Also at: Dipartimento DET del Politecnico di Torino, Turin, Italy\\
$^{\rm VIII}$ Also at: Department of Applied Physics, Aligarh Muslim University, Aligarh, India\\
$^{\rm IX}$ Also at: Institute of Theoretical Physics, University of Wroclaw, Poland\\
$^{\rm X}$ Also at: Facultad de Ciencias, Universidad Nacional Aut\'{o}noma de M\'{e}xico, Mexico City, Mexico\\

\section*{Collaboration Institutes}

$^{1}$ A.I. Alikhanyan National Science Laboratory (Yerevan Physics Institute) Foundation, Yerevan, Armenia\\
$^{2}$ AGH University of Krakow, Cracow, Poland\\
$^{3}$ Bogolyubov Institute for Theoretical Physics, National Academy of Sciences of Ukraine, Kyiv, Ukraine\\
$^{4}$ Bose Institute, Department of Physics  and Centre for Astroparticle Physics and Space Science (CAPSS), Kolkata, India\\
$^{5}$ California Polytechnic State University, San Luis Obispo, California, United States\\
$^{6}$ Central China Normal University, Wuhan, China\\
$^{7}$ Centro de Aplicaciones Tecnol\'{o}gicas y Desarrollo Nuclear (CEADEN), Havana, Cuba\\
$^{8}$ Centro de Investigaci\'{o}n y de Estudios Avanzados (CINVESTAV), Mexico City and M\'{e}rida, Mexico\\
$^{9}$ Chicago State University, Chicago, Illinois, United States\\
$^{10}$ China Nuclear Data Center, China Institute of Atomic Energy, Beijing, China\\
$^{11}$ China University of Geosciences, Wuhan, China\\
$^{12}$ Chungbuk National University, Cheongju, Republic of Korea\\
$^{13}$ Comenius University Bratislava, Faculty of Mathematics, Physics and Informatics, Bratislava, Slovak Republic\\
$^{14}$ Creighton University, Omaha, Nebraska, United States\\
$^{15}$ Department of Physics, Aligarh Muslim University, Aligarh, India\\
$^{16}$ Department of Physics, Pusan National University, Pusan, Republic of Korea\\
$^{17}$ Department of Physics, Sejong University, Seoul, Republic of Korea\\
$^{18}$ Department of Physics, University of California, Berkeley, California, United States\\
$^{19}$ Department of Physics, University of Oslo, Oslo, Norway\\
$^{20}$ Department of Physics and Technology, University of Bergen, Bergen, Norway\\
$^{21}$ Dipartimento di Fisica, Universit\`{a} di Pavia, Pavia, Italy\\
$^{22}$ Dipartimento di Fisica dell'Universit\`{a} and Sezione INFN, Cagliari, Italy\\
$^{23}$ Dipartimento di Fisica dell'Universit\`{a} and Sezione INFN, Trieste, Italy\\
$^{24}$ Dipartimento di Fisica dell'Universit\`{a} and Sezione INFN, Turin, Italy\\
$^{25}$ Dipartimento di Fisica e Astronomia dell'Universit\`{a} and Sezione INFN, Bologna, Italy\\
$^{26}$ Dipartimento di Fisica e Astronomia dell'Universit\`{a} and Sezione INFN, Catania, Italy\\
$^{27}$ Dipartimento di Fisica e Astronomia dell'Universit\`{a} and Sezione INFN, Padova, Italy\\
$^{28}$ Dipartimento di Fisica `E.R.~Caianiello' dell'Universit\`{a} and Gruppo Collegato INFN, Salerno, Italy\\
$^{29}$ Dipartimento DISAT del Politecnico and Sezione INFN, Turin, Italy\\
$^{30}$ Dipartimento di Scienze MIFT, Universit\`{a} di Messina, Messina, Italy\\
$^{31}$ Dipartimento Interateneo di Fisica `M.~Merlin' and Sezione INFN, Bari, Italy\\
$^{32}$ European Organization for Nuclear Research (CERN), Geneva, Switzerland\\
$^{33}$ Faculty of Electrical Engineering, Mechanical Engineering and Naval Architecture, University of Split, Split, Croatia\\
$^{34}$ Faculty of Nuclear Sciences and Physical Engineering, Czech Technical University in Prague, Prague, Czech Republic\\
$^{35}$ Faculty of Physics, Sofia University, Sofia, Bulgaria\\
$^{36}$ Faculty of Science, P.J.~\v{S}af\'{a}rik University, Ko\v{s}ice, Slovak Republic\\
$^{37}$ Faculty of Technology, Environmental and Social Sciences, Bergen, Norway\\
$^{38}$ Frankfurt Institute for Advanced Studies, Johann Wolfgang Goethe-Universit\"{a}t Frankfurt, Frankfurt, Germany\\
$^{39}$ Fudan University, Shanghai, China\\
$^{40}$ Gauhati University, Department of Physics, Guwahati, India\\
$^{41}$ Helmholtz-Institut f\"{u}r Strahlen- und Kernphysik, Rheinische Friedrich-Wilhelms-Universit\"{a}t Bonn, Bonn, Germany\\
$^{42}$ Helsinki Institute of Physics (HIP), Helsinki, Finland\\
$^{43}$ High Energy Physics Group,  Universidad Aut\'{o}noma de Puebla, Puebla, Mexico\\
$^{44}$ Horia Hulubei National Institute of Physics and Nuclear Engineering, Bucharest, Romania\\
$^{45}$ HUN-REN Wigner Research Centre for Physics, Budapest, Hungary\\
$^{46}$ Indian Institute of Technology Bombay (IIT), Mumbai, India\\
$^{47}$ Indian Institute of Technology Indore, Indore, India\\
$^{48}$ INFN, Laboratori Nazionali di Frascati, Frascati, Italy\\
$^{49}$ INFN, Sezione di Bari, Bari, Italy\\
$^{50}$ INFN, Sezione di Bologna, Bologna, Italy\\
$^{51}$ INFN, Sezione di Cagliari, Cagliari, Italy\\
$^{52}$ INFN, Sezione di Catania, Catania, Italy\\
$^{53}$ INFN, Sezione di Padova, Padova, Italy\\
$^{54}$ INFN, Sezione di Pavia, Pavia, Italy\\
$^{55}$ INFN, Sezione di Torino, Turin, Italy\\
$^{56}$ INFN, Sezione di Trieste, Trieste, Italy\\
$^{57}$ Inha University, Incheon, Republic of Korea\\
$^{58}$ Institute for Gravitational and Subatomic Physics (GRASP), Utrecht University/Nikhef, Utrecht, Netherlands\\
$^{59}$ Institute of Experimental Physics, Slovak Academy of Sciences, Ko\v{s}ice, Slovak Republic\\
$^{60}$ Institute of Physics, Homi Bhabha National Institute, Bhubaneswar, India\\
$^{61}$ Institute of Physics of the Czech Academy of Sciences, Prague, Czech Republic\\
$^{62}$ Institute of Space Science (ISS), Bucharest, Romania\\
$^{63}$ Institut f\"{u}r Kernphysik, Johann Wolfgang Goethe-Universit\"{a}t Frankfurt, Frankfurt, Germany\\
$^{64}$ Instituto de Ciencias Nucleares, Universidad Nacional Aut\'{o}noma de M\'{e}xico, Mexico City, Mexico\\
$^{65}$ Instituto de F\'{i}sica, Universidade Federal do Rio Grande do Sul (UFRGS), Porto Alegre, Brazil\\
$^{66}$ Instituto de F\'{\i}sica, Universidad Nacional Aut\'{o}noma de M\'{e}xico, Mexico City, Mexico\\
$^{67}$ iThemba LABS, National Research Foundation, Somerset West, South Africa\\
$^{68}$ Jeonbuk National University, Jeonju, Republic of Korea\\
$^{69}$ Korea Institute of Science and Technology Information, Daejeon, Republic of Korea\\
$^{70}$ Laboratoire de Physique Subatomique et de Cosmologie, Universit\'{e} Grenoble-Alpes, CNRS-IN2P3, Grenoble, France\\
$^{71}$ Lawrence Berkeley National Laboratory, Berkeley, California, United States\\
$^{72}$ Lund University Department of Physics, Division of Particle Physics, Lund, Sweden\\
$^{73}$ Marietta Blau Institute, Vienna, Austria\\
$^{74}$ Nagasaki Institute of Applied Science, Nagasaki, Japan\\
$^{75}$ Nara Women{'}s University (NWU), Nara, Japan\\
$^{76}$ National Centre for Nuclear Research, Warsaw, Poland\\
$^{77}$ National Institute of Science Education and Research, Homi Bhabha National Institute, Jatni, India\\
$^{78}$ National Nuclear Research Center, Baku, Azerbaijan\\
$^{79}$ National Research and Innovation Agency - BRIN, Jakarta, Indonesia\\
$^{80}$ Niels Bohr Institute, University of Copenhagen, Copenhagen, Denmark\\
$^{81}$ Nikhef, National institute for subatomic physics, Amsterdam, Netherlands\\
$^{82}$ Nuclear Physics Group, STFC Daresbury Laboratory, Daresbury, United Kingdom\\
$^{83}$ Nuclear Physics Institute of the Czech Academy of Sciences, Husinec-\v{R}e\v{z}, Czech Republic\\
$^{84}$ Oak Ridge National Laboratory, Oak Ridge, Tennessee, United States\\
$^{85}$ Ohio State University, Columbus, Ohio, United States\\
$^{86}$ Physics department, Faculty of science, University of Zagreb, Zagreb, Croatia\\
$^{87}$ Physics Department, Panjab University, Chandigarh, India\\
$^{88}$ Physics Department, University of Jammu, Jammu, India\\
$^{89}$ Physics Program and International Institute for Sustainability with Knotted Chiral Meta Matter (WPI-SKCM$^{2}$), Hiroshima University, Hiroshima, Japan\\
$^{90}$ Physikalisches Institut, Eberhard-Karls-Universit\"{a}t T\"{u}bingen, T\"{u}bingen, Germany\\
$^{91}$ Physikalisches Institut, Ruprecht-Karls-Universit\"{a}t Heidelberg, Heidelberg, Germany\\
$^{92}$ Physik Department, Technische Universit\"{a}t M\"{u}nchen, Munich, Germany\\
$^{93}$ Politecnico di Bari and Sezione INFN, Bari, Italy\\
$^{94}$ Research Division and ExtreMe Matter Institute EMMI, GSI Helmholtzzentrum f\"ur Schwerionenforschung GmbH, Darmstadt, Germany\\
$^{95}$ Saga University, Saga, Japan\\
$^{96}$ Saha Institute of Nuclear Physics, Homi Bhabha National Institute, Kolkata, India\\
$^{97}$ School of Physics and Astronomy, University of Birmingham, Birmingham, United Kingdom\\
$^{98}$ Secci\'{o}n F\'{\i}sica, Departamento de Ciencias, Pontificia Universidad Cat\'{o}lica del Per\'{u}, Lima, Peru\\
$^{99}$ SUBATECH, IMT Atlantique, Nantes Universit\'{e}, CNRS-IN2P3, Nantes, France\\
$^{100}$ Sungkyunkwan University, Suwon City, Republic of Korea\\
$^{101}$ Suranaree University of Technology, Nakhon Ratchasima, Thailand\\
$^{102}$ Technical University of Ko\v{s}ice, Ko\v{s}ice, Slovak Republic\\
$^{103}$ The Henryk Niewodniczanski Institute of Nuclear Physics, Polish Academy of Sciences, Cracow, Poland\\
$^{104}$ The University of Texas at Austin, Austin, Texas, United States\\
$^{105}$ Universidad Aut\'{o}noma de Sinaloa, Culiac\'{a}n, Mexico\\
$^{106}$ Universidade de S\~{a}o Paulo (USP), S\~{a}o Paulo, Brazil\\
$^{107}$ Universidade Estadual de Campinas (UNICAMP), Campinas, Brazil\\
$^{108}$ Universidade Federal do ABC, Santo Andre, Brazil\\
$^{109}$ Universitatea Nationala de Stiinta si Tehnologie Politehnica Bucuresti, Bucharest, Romania\\
$^{110}$ University of Cape Town, Cape Town, South Africa\\
$^{111}$ University of Derby, Derby, United Kingdom\\
$^{112}$ University of Houston, Houston, Texas, United States\\
$^{113}$ University of Jyv\"{a}skyl\"{a}, Jyv\"{a}skyl\"{a}, Finland\\
$^{114}$ University of Kansas, Lawrence, Kansas, United States\\
$^{115}$ University of Liverpool, Liverpool, United Kingdom\\
$^{116}$ University of Science and Technology of China, Hefei, China\\
$^{117}$ University of Silesia in Katowice, Katowice, Poland\\
$^{118}$ University of South-Eastern Norway, Kongsberg, Norway\\
$^{119}$ University of Tennessee, Knoxville, Tennessee, United States\\
$^{120}$ University of the Witwatersrand, Johannesburg, South Africa\\
$^{121}$ University of Tokyo, Tokyo, Japan\\
$^{122}$ University of Tsukuba, Tsukuba, Japan\\
$^{123}$ Universit\"{a}t M\"{u}nster, Institut f\"{u}r Kernphysik, M\"{u}nster, Germany\\
$^{124}$ Universit\'{e} Clermont Auvergne, CNRS/IN2P3, LPC, Clermont-Ferrand, France\\
$^{125}$ Universit\'{e} de Lyon, CNRS/IN2P3, Institut de Physique des 2 Infinis de Lyon, Lyon, France\\
$^{126}$ Universit\'{e} de Strasbourg, CNRS, IPHC UMR 7178, F-67000 Strasbourg, France, Strasbourg, France\\
$^{127}$ Universit\'{e} Paris-Saclay, Centre d'Etudes de Saclay (CEA), IRFU, D\'{e}partment de Physique Nucl\'{e}aire (DPhN), Saclay, France\\
$^{128}$ Universit\'{e}  Paris-Saclay, CNRS/IN2P3, IJCLab, Orsay, France\\
$^{129}$ Universit\`{a} degli Studi di Foggia, Foggia, Italy\\
$^{130}$ Universit\`{a} del Piemonte Orientale, Vercelli, Italy\\
$^{131}$ Universit\`{a} di Brescia, Brescia, Italy\\
$^{132}$ Variable Energy Cyclotron Centre, Homi Bhabha National Institute, Kolkata, India\\
$^{133}$ Warsaw University of Technology, Warsaw, Poland\\
$^{134}$ Wayne State University, Detroit, Michigan, United States\\
$^{135}$ Yale University, New Haven, Connecticut, United States\\
$^{136}$ Yildiz Technical University, Istanbul, Turkey\\
$^{137}$ Yonsei University, Seoul, Republic of Korea\\
$^{138}$ Affiliated with an institute formerly covered by a cooperation agreement with CERN\\
$^{139}$ Affiliated with an international laboratory covered by a cooperation agreement with CERN.\\

\end{flushleft} 

\end{document}